\documentclass[aip,apl,reprint]{revtex4-1}  
\usepackage{graphicx}
\usepackage{amsfonts}              
\usepackage{amssymb,amsmath}
\usepackage{bm}
\usepackage{array}
\usepackage{natbib}
\usepackage{comment}
\begin{document}
\title{Valley filter in strain engineered graphene}
 \author{T.\ Fujita}       
 \email{t.fujita@nus.edu.sg}
   \affiliation{Computational Nanoelectronics and Nano-device Laboratory, Electrical and Computer Engineering Department, National University of Singapore, 4 Engineering Drive 3, Singapore 117576}
   \author{M.\ B.\ A.\ Jalil}       
   \affiliation{Information Storage Materials Laboratory, Electrical and Computer Engineering Department, National University of Singapore, 4 Engineering Drive 3, Singapore 117576}
   \affiliation{Computational Nanoelectronics and Nano-device Laboratory, Electrical and Computer Engineering Department, National University of Singapore, 4 Engineering Drive 3, Singapore 117576}
    \author{S.\ G.\ Tan}       
    \affiliation{Data Storage Institute, A*STAR (Agency for Science, Technology and Research) DSI Building, 5 Engineering Drive 1, Singapore 117608}
   \affiliation{Computational Nanoelectronics and Nano-device Laboratory, Electrical and Computer Engineering Department, National University of Singapore, 4 Engineering Drive 3, Singapore 117576}
    \date{\today}    

\begin{abstract}
We propose a simple, yet highly efficient and robust device for producing valley polarized current in graphene. The device comprises of two distinct components; a region of uniform uniaxial strain, adjacent to an out-of-plane magnetic barrier configuration formed by patterned ferromagnetic gates. We show that when the amount of strain, magnetic field strength, and Fermi level are properly tuned, the output current can be made to consist of only a single valley contribution. Perfect valley filtering is achievable within experimentally accessible parameters.
\end{abstract}
\maketitle
\section{Introduction}
As an atomically thin sheet of carbon atoms, graphene can be thought of as a flexible membrane. Very recently, the prospect of using strain to engineer the electronic properties of graphene \cite{pereira} has opened up new opportunities and directions for graphene research. Strain essentially can be considered as a perturbation to the in-plane hopping
amplitude, which induces a gauge potential in the effective Hamiltonian.\cite{pereira,sasaki} Previously, strain engineering of graphene has been theoretically applied \emph{e.g.}\ in beam collimation\cite{pereira} and the quantized valley Hall effect.\cite{tlow} Experimentally, controllable strain has been induced in graphene via deposition onto stretchable substrates \cite{ni,tyu} and free suspension across trenches.\cite{lau}

The valleys in graphene refer to the Dirac cones situated at the six corners of the hexagonal Brillouin zone (BZ), of which there are two inequivalent types labeled $K$ and $K'$.\cite{castro} The valley degree of freedom represents a spin-like quantity, and the study of manipulating and making use of valleys in technology has fittingly been termed \emph{valleytronics}.\cite{rycerz} A prerequisite for this technology is a simple and effective method for preparing valley polarized currents. A number of such \emph{valley filters} have been proposed in the literature. \cite{rycerz,abergel,JMPereira_trigonal} The valley filter in Ref.\ \onlinecite{rycerz} operates by passing electronic current through graphene nanoconstrictions. One caveat is that the filter is sensitive to the edge profile of the graphene sample, which is difficult to control practically. The filter in Ref.\ \onlinecite{JMPereira_trigonal}, on the other hand, exploits the valley-asymmetry of trigonal warping which is large for energies far away from the Dirac point. Large energy scales, however, enhance intervalley scattering and adversely affect the filter's efficiency. Lastly, the filter described in Ref.\ \onlinecite{abergel} is effective in bilayer graphene only under intense irradiation by high frequency light.
 
In this letter, we propose an efficient and robust way to filter monolayer graphene's valleys through strain engineering. Significantly, we found that almost pure valley polarized currents can be attained within experimentally relevant parameters. Moreover, it operates in the low energy limit, in which intervalley scattering is negligible.\cite{abergel}
\section{Theory}
\begin{figure}[!ht]
\centering
\resizebox{\columnwidth}{!}{
\includegraphics{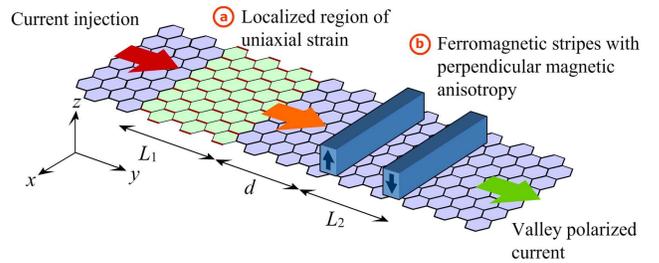}}
\caption{(Color) The proposed valley filter comprising of two cascaded structures; (a) a region of length $L_1$ of uniform uniaxial strain (strained bonds are highlighted in red), and (b) a magnetic barrier structure due to patterned FM top-gates separated by length $L_2$. When the two stages are cascaded, it is shown that the output current is highly polarized in one of graphene's valleys.}
\label{device.fig}
\end{figure}

We consider the five-layered graphene device shown in Fig.\ \ref{device.fig}, comprising of two cascaded components; (a) a region of uniform uniaxial strain, followed by (b) a magnetic field configuration arising from a pair of pattered ferromagnetic (FM) gates. For clarity, the two components are analyzed separately in Sec.\ \ref{strain.sec} and \ref{mag.sec}, respectively. The combined structure is studied in Sec.\ \ref{full.sec}.
\subsection{Valley-dependent tunneling through strained graphene structures}
\label{strain.sec}
\begin{figure}[!ht]
\centering
\resizebox{\columnwidth}{!}{
\includegraphics{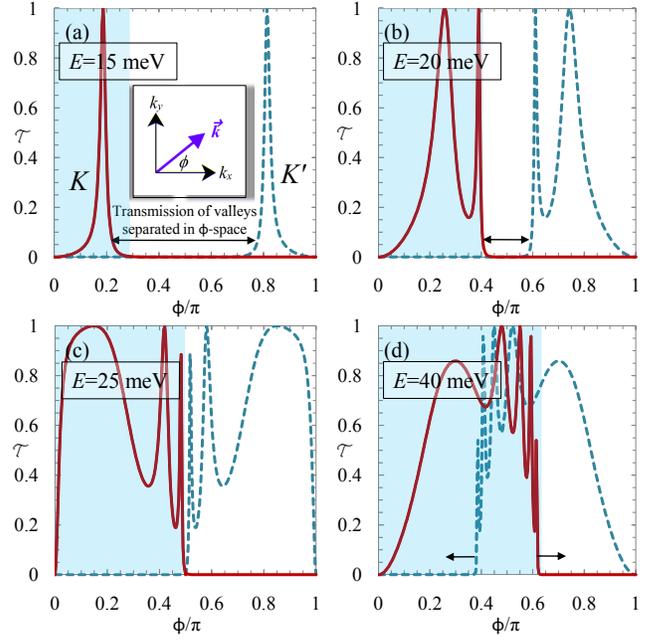}}
\caption{(Color) Transmission probability $\mathcal{T}$ through the strained region in Fig.\ \ref{device.fig}(a) as a function of $\phi$ (see inset of (a)) for various Fermi energies when the strain $\delta t=25$ meV and $L_1=250$ nm. $\mathcal{T}$ carries a strong valley-dependence; for the $K$-valley it is denoted by red, solid curves whilst for the $K'$-valley it is denoted by blue, dashed curves. The shaded blue regions indicate the action (passable values of $\phi$) of a matched $\phi$-filter, which is discussed in Section \ref{mag.sec}.}
\label{valley_dep_trans.fig}
\end{figure}

We assume a region (length: $L_1$) of uniform uniaxial strain along the armchair direction (corresponding to the $\hat{y}$-direction) of monolayer graphene as shown in Fig.\ \ref{device.fig}(a). In the presence of strain, the effective Hamiltonian reads $\mathcal{H}^K=v_F \vec{\sigma}\cdot\left(\vec{p}-v_F^{-1}\vec{A}_S\right)$ for the $K$-valley, and $\mathcal{H}^{K'}=v_F \vec{\sigma}^{'}\cdot\left(\vec{p}+v_F^{-1}\vec{A}_S\right)$ for the $K'$-valley where $\vec{\sigma}^{(')}=\left(\sigma_x,(-)\sigma_y\right)$.\cite{castro} The sign difference between the induced gauge potentials $\vec{A}_S$ ensures overall time-reversal (TR) symmetry. Explicitly, we have $\vec{A}_S=\delta t \hat{x}$,\cite{pereira}
where $\delta t$ parametrizes the strain by its effect on the nearest neighbor hopping, $t\rightarrow t+\delta t$ ($t\approx 3$ eV). To study the transmission across the strained region, we start by writing down the electronic wavefunctions in the three regions. Translational invariance along $\hat{x}$ permits solutions in the $K$-valley of the form $\Psi(x,y)=\exp{(i k_x x)}\Psi(y)$, where
\begin{eqnarray}
\Psi_{\textrm{I}}(y)&=&e^{ik y}\left(\begin{array}{c}1\\e^{i\phi}\end{array}\right)+R e^{-ik y}\left(\begin{array}{c}1\\e^{-i\phi}\end{array}\right),\nonumber\\
\Psi_{\textrm{II}}(y)&=&A e^{iq y}\left(\begin{array}{c}1\\e^{i\varphi}\end{array}\right)+B e^{-iq y}\left(\begin{array}{c}1\\e^{-i\varphi}\end{array}\right),\nonumber\\
\Psi_{\textrm{III}}(y)&=&Te^{ik y}\left(\begin{array}{c}1\\e^{i\phi}\end{array}\right),\nonumber
\end{eqnarray}
and $k_x^2+k^2=E^2=(k_x\mp\delta t)^2+q^2$. For the $K'$-valley, we replace $\phi\rightarrow -\phi$, $\varphi\rightarrow -\varphi$. The phases $\phi$ and $\varphi$ are parametrized as $k_x = E\cos\phi$ and $k_x\mp\delta t=E\cos\varphi$, where the upper (lower) sign corresponds to valley $K$ ($K'$), and are related via the conservation of $k_x$; $E\cos\phi = E\cos\varphi\pm\delta t$ (in this paper we assume $\delta t >0$ and $E>0$). The transmission coefficient $T$ is solved via wavefunction continuity, and determines the transmission probability $\mathcal{T}=|T|^2$. In Fig.\ \ref{valley_dep_trans.fig} we plot $\mathcal{T}$ as a function of $\phi$ [assuming $\delta t=25$ meV and $L_1=250$ nm], which reveals a remarkable valley-dependence.\cite{voz} In particular, we find that transmission of opposite valleys is well separated in $\phi$-space for low energies $E\le \delta t$. 
The behavior of $\mathcal{T}$ can be understood from the following. From $k_x^2+k^2=E^2$ we require $|k_x|\le E$ for non-vanishing transmission. Similarly, from $(k_x\mp\delta t)^2+q^2=E^2$ we get $-E\pm\delta t\le k_x\le E\pm\delta t$. Combining the two inequalities, $\phi$ must satisfy $-1+\delta t/E\le\cos\phi\le 1$ in the $K$-valley and $-1\le\cos\phi\le 1-\delta t/E$ in the $K'$-valley. Thus, electron transmission is restricted to valley-dependent windows of
\begin{equation}
\phi \in \left\{
\begin{array}{ll}
\left[0,\arccos(-1+\delta t/E) \right]; & K \\
\left[\arccos(1-\delta t/E),\pi \right]; & K' \\
\end{array}
\right. .
\label{phi.eq}
\end{equation}
Evidently, when $\delta t = 0$ (no strain), electron transmission spans the entire spectrum of $\phi\in[0,\pi]$ for both valleys. The onset of total separation occurs at $E=\delta t$ \eqref{phi.eq}, as confirmed in Fig.\ \ref{valley_dep_trans.fig}(c). This separation of the two valleys in $\phi$-space forms the basis of our proposed valley filter. In the next section we design a simple structure which filters electrons in $\phi$-space.
\subsection{$\phi$-space filtering via magnetic fields}
\label{mag.sec}
The required $\phi$-filter is hinted from our analysis
above. Fig.\ \ref{valley_dep_trans.fig} shows that each of the valleys undergo separate $\phi$-space filtering. The fact that the filtering characteristics are mirrored about $\phi = \pi/2$ (equivalent
to $k_x\rightarrow -k_x$) reflects the TR symmetry. For the current application, we require a gauge potential $\vec{A}$ which breaks TR, \emph{i.e.}\ a magnetic field. In Sec.\ \ref{strain.sec}, $\vec{A}_S$ assumed a
rectangular function in the region of uniaxial strain; this can be matched by a magnetic vector potential $\vec{A}$ due to a pair of asymmetric, magnetic
$\delta$-barriers which can be formed via patterned FM gating structures (see Fig.\ \ref{device.fig}(b)).\cite{majumdar} For a magnetic field configuration $B_z(y) = B_0 l_B \left[\delta(y) - \delta(y - L_2)\right]$,
where $l_B = \sqrt{\hbar/|e|B_0}$ and $e < 0$, the TR-breaking gauge
potential reads $\vec{A} = B_0 l_B \left[\Theta(y) -\Theta(y - L_2)\right]\hat{x}$ and enters the Hamiltonian in a valley-independent manner as $\mathcal{H} = v_F \vec{\sigma}^{(')}\cdot\left(\vec{p}+e\vec{A}\right)$. Transmission across
the magnetic barrier is identical to the red, solid curves in
Fig.\ \ref{valley_dep_trans.fig} for matched values $v_F^{-1}\delta t = |e|B_0 l_B$ corresponding
to a field strength of
\begin{equation}
B_0 = \frac{1}{\hbar |e|}\left( \frac{\delta t}{v_F}\right)^2.
\label{matchB.eq}
\end{equation}
A strain of $\delta t = 25$ meV, for example, requires a field strength
of $B_0 = 1$ T. The pass band of the filter when $B_0$ is matched to the
strain \eqref{matchB.eq} is shown in Fig.\ \ref{valley_dep_trans.fig} as the blue, shaded regions.
Perfect valley filtering can then be expected so long as
$E_F \le \delta t$, \emph{i.e.}\ there is no mixing of valley currents in $\phi$-space. Before proceeding, we note that the wavevector filtering property of magnetic fields is generic,\cite{matulis} and not limited to $\delta$-type barriers, which are chosen
above for simplicity. Wavevector filtering occurs just as effectively in graphene for uniform magnetic fields and other periodic
structures.\cite{masir2,anna,ghosh}
\subsection{Valley filtering in cascaded structure}
\label{full.sec}
We analyzed the combined structure in Fig.\ \ref{device.fig}, confirming
our predictions for fully valley polarized currents. The
conductance of the device is
\begin{equation}
G^{K(K')}=G_0 \int_0^\pi d\phi \sin\phi\mathcal{T}^{K(K')}\left(E_F,E_F\cos\phi\right),
\end{equation}
where $E_F$ is the Fermi energy, $\mathcal{T}(E,k_x)$ is the valley-dependent
transmission probability and $G_0 = 2 e^2/h \left( E_F L_x /\hbar v_F \right)$ [$L_x$ is the width along $\hat{x}$]. We define the
\emph{valley polarization} as
\begin{equation}
\mathcal{P} = \frac{G^K - G^{K'}}{G^K + G^{K'}},
\end{equation}
where $|\mathcal{P}|\le 1$. Assuming a device with matched stages \eqref{matchB.eq} and the following parameters; $E_F = \delta t = 25$ meV, $B_0 = 1$ T, and
$L_1 = d = L_2 = 4l_B \approx 100$ nm, we obtain for the
conductances $G^K = 0.310 G_0$ and $G^{K'}
= 1.31 \times 10^{-3} G_0$,
which results in $\mathcal{P} > 0.99$. These values represent experimentally relevant conditions; magnetic field barriers of $B_0\sim 1$ T may be systematically fabricated at these length scales,\cite{ghosh} and strains in graphene of $\approx 1\%$  have already been achieved \cite{ni} ($\delta t = 25$ meV corresponds to $\approx 0.8\%$ strain). 
\section{Analysis and discussion}
\begin{figure}[!ht]
\centering
\begin{tabular}{m{0.1\columnwidth}m{0.85\columnwidth}}
(a) &
\resizebox{0.7\columnwidth}{!}{
\includegraphics{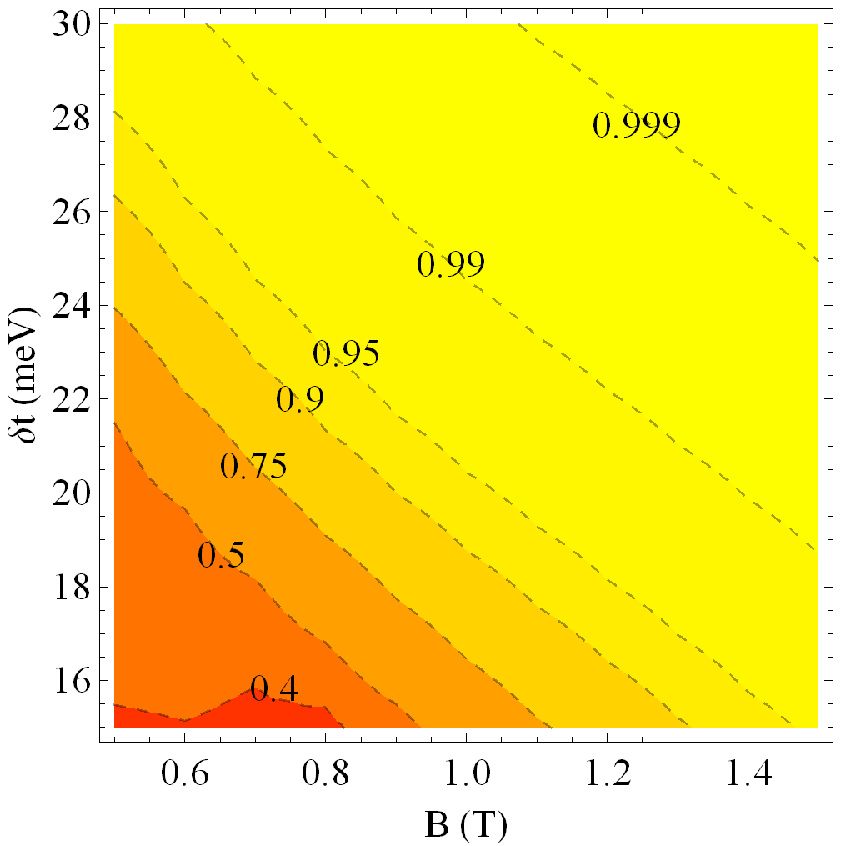}} \\
(b) &
\resizebox{0.7\columnwidth}{!}{
\includegraphics{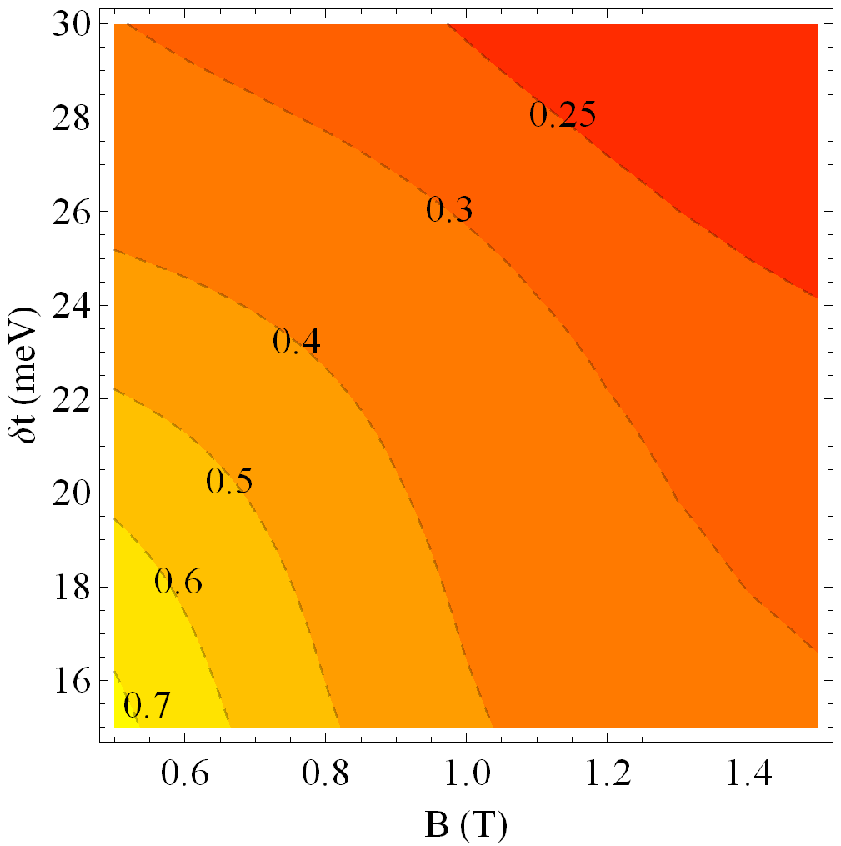}}
\end{tabular}
\caption{(Color) Variation of (a) valley polarization $\mathcal{P}$ and (b) $K$-valley electronic conductance $G^K/G_0$ of the device shown in Fig.\ \ref{device.fig}, for various values of strain $\delta t$ and magnetic field strength $B_0$. Parameter values used are Fermi energy $E_F=25$ meV and lengths $L_1=d=L_2\approx 100$ nm.}
\label{P.fig}
\end{figure}

We analyze the robustness of $\mathcal{P}$ for \emph{unmatched} filter
stages keeping the Fermi level constant. In Fig.\ \ref{P.fig}(a), we plot $\mathcal{P}$
for continuous values of the strain $\delta t \in [15, 30]$ meV and
magnetic field strength $B_0 \in [0.5, 1.5]$ T, for fixed $E_F=25$ meV and $L_1 = d = L_2 \approx 100$ nm. 
%
As is evident, $\mathcal{P}$ remains large and is robust ($> 0.99$) for
a large window of strain and magnetic field values. This is coupled with a significant conductance $G^K$ 
(normalized to $G_0$) as shown in Fig.\ \ref{P.fig}(b). We therefore expect the proposed valley filter to be relevant for practical valleytronic applications which demand a very high and robust valley polarization \emph{and} an appreciable electronic current. The polarity of the filter ($K \rightarrow K'$) can be interchanged in one of two ways. Firstly, one could reverse
the direction of the magnetic field $B_z \rightarrow -B_z$, which
flips the effect of the $\phi$-filter. Alternatively, one may reverse the strain ($\delta t \rightarrow -\delta t$), which interchanges the curves for $K$ and $K'$ in Fig.\ \ref{valley_dep_trans.fig}. 
Finally, it has been shown that strain can induce a bandgap in graphene. However, this occurs for large strains of $\sim 10$--$20\%$ and is due to the shifting of the valleys away from the BZ corners.\cite{pereira-ribeiro} For the small strains used here, we assume a negligible deformation of the valleys.
%

\end{document}